\documentclass[runningheads]{llncs}

\usepackage[T1]{fontenc}
\usepackage{graphicx}
\usepackage{microtype}
\usepackage{amsmath}
\usepackage[hidelinks]{hyperref}

\usepackage[dvipsnames]{xcolor}
\usepackage[utf8]{inputenc}
\usepackage{amsfonts}

\usepackage{longtable}
\usepackage{multirow}
\usepackage{booktabs}
\usepackage{float}
\usepackage{algorithm}
\usepackage{algpseudocode}
\usepackage{paralist}
\usepackage{subcaption}
\usepackage{pbox}
\usepackage{makecell}
\usepackage{cleveref}

\usepackage[
  backend=biber,
  doi=false,isbn=false,url=true,
  maxbibnames=4
]{biblatex}

\begin{document}

\title{Determining Window Sizes using Species Estimation for Accurate Process Mining over Streams}

\author{
Christian Imenkamp\inst{1}        \orcidID{0009-0007-4295-1268} \and
Martin Kabierski\inst{2}     \orcidID{0000-0002-9852-7489} \and
Hendrik Reiter\inst{3}        \orcidID{0009-0003-8544-0012} \and
Matthias Weidlich\inst{2}        \orcidID{0000-0003-3325-7227} \and
Wilhelm Hasselbring\inst{3}        \orcidID{0000-0001-6625-4335} \and
Agnes Koschmider\inst{1,2}        \orcidID{0000-0001-8206-7636}
}

\authorrunning{Imenkamp et al.}
\titlerunning{Determining Window Sizes using Species Estimation}

\institute{Business Informatics and Process Analytics, University of Bayreuth, Germany\\
\email{ \{christian.imenkamp|agnes.koschmider\}@uni-bayreuth.de}\\
\url{https://www.pa.uni-bayreuth.de/en/}
\and
Department of Computer Science, Humboldt-Universität zu Berlin, Germany\\
\email{ \{martin.kabierski|matthias.weidlich\}@hu-berlin.de}\\
\and
Department of Computer Science, Christian-Albrechts-Universität zu Kiel\\
\email{ \{hendrik.reiter|hasselbring\}@email.uni-kiel.de}\\
}

\maketitle             

\begin{abstract}

Streaming process mining deals with the real-time analysis of event streams. A common approach for it is to adopt windowing mechanisms that select event data from a stream for subsequent analysis. However, the size of these windows denotes a crucial parameter, as it influences the representativeness of the window content and, by extension, of the analysis results. Given that process dynamics are subject to changes and potential concept drift, a static, fixed window size leads to inaccurate representations that introduce bias in the analysis.
In this work, we present a novel approach for streaming process mining that addresses these limitations by adjusting window sizes. Specifically, we dynamically determine suitable window sizes based on estimators for the representativeness of samples as developed for species estimation in biodiversity research. Evaluation results on real-world data sets show improvements over existing approaches that adopt static window sizes in terms of accuracy and robustness to concept drifts.

\end{abstract}

\section{Introduction}
Generally, process mining techniques may be employed in \textit{offline} and \textit{online} use cases. Techniques for offline process mining evaluate historical event data, while streaming process mining aims at providing immediate insights based on a continuous stream of event data. As such, streaming techniques handle event data as soon as it becomes available, thereby enabling a timely reaction to unexpected process dynamics. Streaming techniques for process mining may generally adopt three different paradigms~\cite{DBLP:books/sp/22/Burattin22}: (i) Window-based approaches select a particular part of the event data at specific time points~\cite{DBLP:books/sp/22/Burattin22}, which then facilitates the analysis using traditional process mining techniques. (ii) Other approaches implement a problem reduction, i.e., they trace back the computations required for streaming process mining to more generic streaming problems~\cite{DBLP:books/sp/22/Burattin22}, such as counting or pattern detection over data streams. (iii) Finally, streaming process mining may rely on the pre-computation of possible analysis results, which are then merely selected based on the event data encountered in a stream~\cite{DBLP:books/sp/22/Burattin22}.

Considering these paradigms, window-based approaches have the major advantage that they do not impose any assumption on the analysis algorithms to use. Unlike the approaches that rely on traditional streaming algorithms or pre-computation, they provide a generic solution for a wide range of specific analysis algorithms. However, when adopting a window-based approach, an important challenge is the selection of suitable parameters for the applied windows. While the question of the \emph{points in time} for the evaluation of a window (commonly referred to as the window \emph{slide}) is typically answered directly based on the latency requirements of the analysis, the question of a suitable \emph{size} of the window imposes important challenges. While the window size should be selected so that the window is \emph{representative} of the underlying process dynamics (i.e., how well does a sample K (the content of the window) captures or preserve important characteristics of a population P (a generative system or the origin of the event stream)). In practice, processes and hence the event data generated by them tend to show certain variability, stemming from exceptional process executions or concept drift. A fixed, static size of a window can, therefore, not be expected to capture a process accurately, which biases the downstream analysis tasks.

Let's consider the following example to motivate the importance of the optimal window size: A hospital monitors thousands of patient treatment processes daily, utilizing an information system. The hospital can therefore derive information on process conformance, staff utilization and treatment protocols in real-time. However, the treatment processes naturally evolve due to factors like seasonal changes, staff rotation or new medical protocols. It's likely that a static window size will result in: (1) It will miss important changes during high-variation periods if the window size is too small. (2) The window include outdated information and therefore decrease the accuracy if the window is too big. Moreover, the higher window size will result in higher resource consumption.

In this paper, we address the challenge of selecting representative windows for streaming process mining. Our approach builds upon recent results on assessing the representativeness of event data for process mining using notions and measures from biodiversity research, particularly relying on species estimation. Specifically, we contribute an approach for the dynamic adaptation of the size of a window, based on its estimated representativeness of the underlying process, in terms of the completeness and coverage of the species induced by the event data. Additionally, we provide an algorithm to automatically adjust the coverage threshold of the window. We evaluate our approach in three steps. First, we demonstrate robustness to concept drift by using synthetic event logs with different drift types. We show how the window automatically adapts to the new stream characteristics. Next, we assess the accuracy of the windows in terms of the F1-score of the discovered process model utilizing real-world event logs and benchmark our approach against existing windowing methods. Finally, we show the real-world applicability by analyzing the runtime efficiency and latency.

In the remainder, \Cref{sec:background} provides the background for our work. \Cref{speciesdiscovery} presents our approach for the selection of a window size. \Cref{evaluation} discusses evaluation results, \Cref{sec:related_work} reviews related work, and \Cref{sec:conclusion} presents conclusions.


\section{Background}
\label{sec:background}
This section introduces preliminaries for our work, in terms of representativeness, an event stream model~(\Cref{sec:event_stream}), sliding windows~(\Cref{sec:window_size}), concept drifts~(\Cref{subsec:concept_drifts}) and estimation techniques for the completeness and coverage of event data~(\Cref{sec:background_completeness_estimation}).

\subsection{Event Stream Model}
\label{sec:event_stream}
Following~\cite{burattin_online_2015}, we define an event as a triple $e=(c, a, t)$, where $c$ is a case identifier, $a$ is an activity name, and $t$ is a timestamp. We write $e.c$, $e.a$, and $e.t$ to refer to the components of event $e$. The set of all possible events is called the event universe, denoted as $\mathcal{E} = \mathcal{C} \times \mathcal{A} \times \mathbb{N}$, where $\mathcal{C}$ is the set of all case identifiers and $\mathcal{A}$ is the set of all activity names. A potentially infinite sequence of events $S: \mathbb{N} \to \mathcal{E}$ is called an event stream. Here, we assume that the order of events in the stream complies with their temporal order, i.e., for all $i,j \in \mathbb{N}$,  it holds that $i<j$ implies that $S(i).t <S(j).t$.

\subsection{Windowing Methods}
\label{sec:window_size}
To process a stream of event data, it is common to employ sliding windows that characterize the elements that are currently active. We shortly review prominent notions of sliding windows that are used in the context of streaming process mining~\cite{burattin_streaming_2022}. 
In a time-based window, events are gathered within a predefined time interval. For count-based windows, both the size and slide parameters are given in terms of a fixed number of events. In a landmark window, a designated event serves as a ``landmark'' that indicates the end of a window.

\subsection{Concept Drifts}
\label{subsec:concept_drifts}
Concept drift occurs when the behavior of a process changes over time in unforeseen ways \cite{schlimmer_beyond_1986}. Drifts can occur in many different ways (i) \textbf{Sudden Drift}: The behavior of the stream changes abruptly. (ii) \textbf{Recurring Drift}: Recurring changes appear seasonally over time. (iii) \textbf{Gradual Drift}: Gradual changes occur through slow degradation, starting in specific contexts and eventually affecting the entire stream. (iv) \textbf{Incremental Drift}:  Incremental changes involve many small-scale modifications, leading to changed behavior over time.


\subsection{Completeness and Coverage Estimation}
\label{sec:background_completeness_estimation}

To assess the representativeness of event data, we adopt measures for the completeness and coverage of event logs~\cite{kabierski_addressing_2023}. They relate the number of observed distinct values of interest in an event log, called \emph{log species}, to the estimated total number of species, the \emph{species richness}, in the system generating the event log.

We here define auxiliary concepts, as follows. Let $P$ be a potentially infinite population of individuals, each belonging to exactly one species, whose occurrence probabilities are fixed, unknown and assumed to be independent from each other. Let $\Omega_P={1,2,\dots,S_P}$ be the finite, enumerated set of species in $P$, i.e. \emph{the species richness} of $P$ with each species having a \emph{species occurrence probability} of $p_1,p_2,\dots,p_{S_P}$, respectively. Furthermore, let $\zeta:P\rightarrow {\Omega_P}$ be a \emph{species retrieval function}, which assigns to an individual, the species it belongs to.

Now, consider a sample $N \subseteq P$ of size $n=|N|$ that is drawn
randomly from the possibly infinite population $P$. By $S_N\leq S_P$, we denote the number of species observed in sample $N$. Let $X_i$ be the \emph{sample species frequency} of species $i$, i.e. the total number of occurrences of species $i$ in sample $N$, so that any species with $X_i=0$ is unobserved in sample $N$.

Based thereon, $f_k=\sum_{i=1}^{S_P} I(X_i=k)$ is the \emph{abundance frequency count}, i.e., the number of species that are represented by exactly $k$ observations in $N$. We denote as $f_1$ the number of species represented by exactly one individual
(called a \emph{singleton}) and as $f_2$ the number of species denoted by exactly two individuals (called a \emph{doubleton}). Under this multinomial sampling model of abundance-based species data~\cite{colwell_models_2012}, the probability of observing species counts $X_1=x_1,X_2=x_2,\dots,X_{S_P}=x_{s_P}$ is given as:
\begin{equation}
\begin{split}
P(X_1=x_1,X_2=x_2,\dots,X_{S_P}=x_{s_P})&= \\
\frac{n!}{x_1!x_2!\dots x_{s_P}!}&p_1^{x_1}p_2^{x_2}\dots p_{S_P}^{x_{s_P}}.
\end{split}
\end{equation}
For this data model, we obtain an estimate of the species richness $S_P$ using the Chao1-estimator~\cite{chao_nonparametric_1984} as:
\begin{equation}
    	\hat{S}_{Chao1} \approx \begin{cases}
		S_{\mathit{N}} + f_1^2/\left(2f_2\right) & \text{if }f_2 > 0,\\
		S_{\mathit{N}} + f_1\left(f_1-1\right)/2 & \text{if }f_2 = 0.
	\end{cases}
\label{eq:chao1}
\end{equation}

Using the estimated species richness $\hat{S}_{Chao1}$, we then derive the \emph{completeness} of sample $N$, i.e. the fraction of observed species from $P$ in $N$, denoted $\hat{Com_{N}}$, as:
\begin{equation}
    \hat{Com_{N}} = \frac{S_{N}}{\hat{S}_{Chao1}}
\label{eq:completeness}
\end{equation}
Furthermore, we obtain the \emph{coverage} of sample $N$, i.e. the probability space covered by by the observed species in $N$, as:
\begin{equation}
    \hat{Cov_{N}} = 1 - \frac{f_1}{n}\left(1-\frac{2f_2}{(n-1)f_1+2f_2}\right)
    \label{eq:coverage}
\end{equation}

The Chao1-estimator yields a lower bound on $S_P$, and for the optimal case, in which the occurrence probabilities of all undetected species are equal, it is an unbiased point estimator of $S_P$~\cite{chao_thirty_2017}. 

For samples, for which $f_1=0$, the right-hand side of \Cref{eq:chao1} evaluates to $0$, which indicates the sample to have full completeness and coverage.

We note, that in general, the above metrics are subject to various assumptions and are thus suspectible to certain scenarios in which these do not hold.
First, the estimators assume that each species will eventually be reobserved under infinite sampling. Clearly, should this not be the case, for instance in the case of erroneous events, that are recorded only once, in the case of noisy event streams, which may also generate spurious events, or in the case of process drifts, that remove previously existent behavior, then perfect completeness may never be reached. Coverage on the other hand will still converge to full coverage, due to the diminishing impact of these spurious species for increasingly growing samples. Furthermore, the species domain needs to be finite, i.e., the number of species in the system needs to be bounded, as estimates will otherwise approach infinity. 


For a small sample, in turn, it may occur that no singleton species has been sampled, which inadvertently indicates full completeness and coverage. To avoid this effect, the sample size shall be increased, as to ensure a proper estimate of the species richness, see also~\cite{kabierski_addressing_2023}.

In contrast to ~\cite{kabierski_addressing_2023}, we here utilize the Chao1-estimator, instead of the Chao2-estimator. While similar in structure, Chao1 assumes each observation to consist of exactly one species and retains exact occurrence counts per species, whereas Chao2 assumes observations to contain multiple species and relies on a different sampling model that does not account for occurrences of the same species in one observation~\cite{colwell_models_2012}. Since event streams consist of individual observations containing atomic information, we thus opt for estimation using the Chao1-estimator.


Lastly, we note that, while the multinomial sampling model assumes independence of species, estimates are still accurate in the case of dependence between species, as long as the occurrence counts of each species are approximately proportional to the global occurrence probabilities in the population.

\subsection{Species Definitions}
\label{subsec:species_def}
In process analysis, we categorize process characteristics as distinct species, for which the above metrics are computed. Intuitively, each observed species represents an abstraction of process behavior observed in the event stream, with the type of abstraction defining the species view of interest. We define $\zeta:P\rightarrow {\Omega_P}$ as a species retrieval function that assigns to each individual process element its species. Following~\cite{kabierski_addressing_2023}, we utilize three categories of species definitions: Simple Species - Activities ($\zeta_{act}$), Complex Species - Trace Variants ($\zeta_{tv}$) and Intermediate Species - Directly Follows Relations ($\zeta_{df}$). (1) $\zeta_{act}$ defines each activity of a trace as an observed species, i.e. $\Omega_P$ is the set of activities in the system. 
(2) $\zeta_{tv}$ treats each trace as the occurrence of a species, i.e. $\Omega_P$ is the set of trace variants. (3) $\zeta_{df}$ represents the directly follows relations between activities as species, i.e. $\Omega_P$ is the set of directly-follows relations in the system.

\section{Dynamic Window Size Selection}
\label{speciesdiscovery}
We now turn to the discussion of the representativeness of streaming windows. First, in \Cref{subsec:completeness_motivation}, we motivate, why completeness and coverage should be considered as a measure of representativeness, when selecting a window size in streaming process mining. Then, we propose a dynamic threshold heuristic to determine the respective threshold for the coverage of a window \Cref{subsec:finding_the_accurate_window_size}, before explaining, how completeness and coverage may guide the selection of an appropriate windows size in~\Cref{subsec:completeness_streaming}.

\subsection{Motivation for Window Representativeness}
\label{subsec:completeness_motivation}
When aiming at process discovery based on a stream of events, the selection of a window size needs to consider the variability of the data, in addition to its volume and velocity. Specifically, data variability is manifested in the number of distinct activities for which execution is indicated by the events in the stream, and by extension, the behavioral dependencies between activities, which usually build the foundation for state-of-the-art process mining algorithms. Here, missing a crucial activity or dependency may induce significant and unpredictable changes in analysis results. Hence, the window size shall be sufficiently large to provide a \emph{representative} view of the activities and their dependencies regarding the process underlying the stream.

Intuitively, if a process is rather homogenous, i.e. the number of activities and their dependencies are relatively small and stretched out over a short interval, a time-based or count-based windows of small size is sufficient for capturing the process' dynamics. Conversely, for a rather heterogeneous process, with many activities and dependencies between them that are observed over a long time, we expect larger window sizes for capturing all possible behaviors. Yet, process dynamics can be expected to change between these extreme poles, being subject to concept drift. With the event stream becoming more homogenous or heterogeneous as a result of these drifts, the window size shall, therefore, decrease or increase accordingly, to accurately capture the process.
%

We illustrate the above intuition with two simple event streams $S_1$ consisting of events indicating the continuous repeated execution of activities $\langle A,B,C \rangle$ in this order, and $S_2$, consisting of events indicating the continuous and repeated execution of activities $\langle A,B,C,D,E \rangle$, in the same order as well. Events are introduced to both streams with the same frequency. Assume that we want to select a window size, aiming at a representative view of the process' behavior, in terms of completeness.

In the light of the repeating patterns in the streams, we could conclude the view on the stream to be complete sooner for $S_1$ than for $S_2$, since activities start reappearing earlier, at the fourth event, compared to $S_2$, where repetition materializes with the sixth event. Thus, independently of when exactly we are confident that we have seen enough evidence for a complete collection of events, we conclude that the window size for $S_2$ needs to be larger than the window size for $S_1$, based only on the number of observed activities and their occurrence counts in the stream. Likewise, if $S_1$ drifts into $S_2$ or the other way around, then the window size could be increased or decreased according to the observed change and repetition of behavior.

\subsection{Selecting a Window Size}

\label{subsec:finding_the_accurate_window_size}
Any open window should be closed, once the species observed in the windows lifetime exceed a dynamically chosen completeness threshold.
For this selection, the coverage (see \Cref{eq:coverage}) is preferred over completeness (see \Cref{eq:completeness}) in scenarios with drifts or errors, as the former still converges to a measure of completeness in the presence of spurious singleton species.
Here, we introduce a dynamic threshold heuristic to determine this threshold (see \Cref{alg:dyThHr}).  The algorithm is inspired by the elbow method~\cite{8549751} also used for threshold heuristics in~\cite{10.1007/978-3-319-98648-7_16}. First, the algorithm approximates the second derivative, i.e., second-order differences (lines 3-6). In particular, it identifies the point at which additional increases in window size provide diminishing returns. To ensure that open windows are eventually closed, the algorithm furthermore checks for stagnation in completeness (lines 8-15). If stagnation is detected (line 8), it increases a smoothing factor and consequently decreases the window size (lines 9 \& 10). Otherwise, this smoothing factor is reduced (lines 12 \& 13), allowing for slower changes (i.e., to avoid rapid adjustments) in the window size. This ensures that the window size does not fluctuate too quickly in response to short-term changes in coverage. The algorithm assumes responsibility for establishing the threshold, which consequently affects the window size. However, new parameters (e.g., smoothing factors, decay rate) are introduced. Nevertheless, these do not require alteration or adaptation to accommodate new data sets, and a (start) smoothing factor of 0.2 and decay rate of 0.1 can be employed universally.

\begin{algorithm}
\caption{Dynamic Threshold Heuristic}
\label{alg:dynamic_threshold_heuristic}
\begin{algorithmic}[1]

\State \textbf{Input:} coverage history $C_i$,|$C_i$| n, smoothing factor $\text{sf}$, current threshold $\text{ct}$, decay rate $\text{dr}$, minimum threshold $\text{mt}$, stagnation threshold \textbf{$\delta$}, stagnation window $w$, 
\State \textbf{Output:} Updated threshold $\text{ct}_{\text{new}}$

\Comment{Compute second-order differences (a discrete approximation of the second derivative)}
\For{$i = 1$ \textbf{to} $n - 2$}
    \State $r''(i) \gets C_{i-1} - 2C_i + C_{i+1}$ 
\EndFor

\State $i^* \gets \underset{i}{\arg\max} \left( r''(i) \right)$
\State $C_{\text{optimal}} \gets C_{i^* + 1}$ \Comment{Optimal threshold value (i.e., the "elbow" point)}

\Comment{Check for coverage stagnation}
    
\If{$n \geq w$ \textbf{and} $\left| C_k - C_{k+1} \right| < \delta \ \forall k = n-w \textbf{ to } n-2$}
    \State $\text{sf} \gets \min(1.2 \times \text{sf}, \ 0.99)$ \Comment{Increase smoothing}
    \State $\text{ct}_{\text{temp}} \gets \max(\text{ct} - \text{dr}, \ \text{mt})$ \Comment{Decrease threshold}
\Else
    \State $\text{sf} \gets \max(0.8 \times \text{sf}, \ 0.01)$ \Comment{Decrease smoothing for faster reaction}
    \State $\text{ct}_{\text{temp}} \gets \text{ct}$ \Comment{Threshold remains unchanged}
\EndIf

\Comment{Smoothly update the threshold}
\State $\text{ct}_{\text{new}} \gets \text{sf} \times C_{\text{optimal}} + (1 - \text{sf}) \times \text{ct}_{\text{temp}}$

\State \Return $\text{ct}_{\text{new}}$

\end{algorithmic}
\label{alg:dyThHr}
\end{algorithm}

\subsection{Estimations for Streaming Process Discovery}
\label{subsec:completeness_streaming}

To assess the representativeness of a window in the context of streaming process discovery, we operationalize the above estimators as follows. Given a window containing the events $e_1, e_2,\dots, e_n$, we treat each of the $n$ events as an observed individual. Based on the ordered list of events $e_1, e_2, \dots, e_n$, we obtain an ordered list of associated activities as $e_1.act, e_2.act, \dots, e_n.act$. Using this list,we may instantiate any of the species definitions discussed in \Cref{subsec:species_def}.


Consider a sliding window containing events that yield the following sequence of activities $\langle A,B,A,C,B,D,A,C,E\rangle$. Considering $\zeta_{act}$, species $A$ occurs three times, species $B$ and $C$ two times, and species $D$ and $E$ once. Thus, there are five species observed in the sample, $S_{N}=5$, and the abundance frequency counts are $f_3=1$, $f_2=2$, and $f_1=2$. Here, an estimate of the species richness gives $\hat{S}{Chao1}=5 + \frac{2^2}{2\cdot 2} = 6$. That is, based on the seen species, we expect six distinct activities to be present in the process, which yields estimates of $\hat{Com}{N}=\frac{5}{6}\approx0.83$ and $\hat{Cov}_{N}=1-\frac{2}{9}\left(1-\frac{2\cdot 2}{(9-1)2+2\cdot 2}\right)\approx0.82$. By considering, $\zeta_{df}$ we obtain the species $AB,BA,AC,CB,BD,DA,CE$ with corresponding occurrence counts $1,1,2,1,1,1,1$, respectively.

In a streaming setting, the estimators need to be evaluated continuously and efficiently upon the arrival of new events. Assuming that all relevant data structures are initialized before a new event $e$ is introduced into the stream, the algorithmic complexity of the estimation depends on three steps: (1) Retrieving the species from $e$, (2) Updating the observed species counts, (3) Updating the completeness and coverage measures.


The species retrieval function $\zeta$ can be computed in $O(1)$ for a given event $e$. Since this yields a single activity associated with $e$, updating the species counts can be done in $O(S_N)$, assuming a hash-based implementation of the species counts. Lastly, updating the completeness and coverage measures requires the retrieval of $f_1$, $f_2$, $n$, and $S_n$ from the updated species counts. The value of $n$ is obtained in $O(1)$ by incrementing a counter upon arrival of a new event. $S_n$ can be obtained in $O(1)$ from the species counts.

$f_1$ and $f_2$ can be computed without iterating over all observed species for each event by updating a counter for $f_1$ and $f_2$ as follows. Should the species count after updating be $1$, i.e., the current species created by event $e$ is a newly observed singleton, increment $f_1$. Should the updated count be $2$, i.e., a doubleton species, decrement the counter for $f_1$ and increment the counter for $f_2$. Lastly, should the updated count be $3$, decrement the counter for $f_2$. Since the updates are tied to the update of the species counts, they can be computed in $O(1)$. Based on $f_1$, $f_2$, $n$, and $S_N$, we obtain $\hat{S}{Chao1}$, $\hat{Com}{N}$, and $\hat{Cov}_{N}$ in $O(1)$. Thus, the overall time complexity of updating the estimators is $O(S_N)$.

\section{Evaluation Results}
\label{evaluation}
This section summarizes results of evaluations of different aspects of the windowing method. First, we describe the experimental setup (\Cref{sec:eval_setup}). We then report on our results regarding the impact of process drifts (\Cref{subsec:robustness_to_concept_drifts}), the role of species definitions on the accuracy (\Cref{subsec:accuracy}) and the performance of the method (\Cref{subsec:computational_efficency}).

\subsection{Experimental Setup}
\label{sec:eval_setup}
We implemented a prototype in Python\footnote{\href{https://github.com/chimenkamp/adaptive-window-for-online-process-mining}{https://github.com/chimenkamp/adaptive-window-for-online-process-mining}}. The implementation is based on processing events that are emitted via the distributed stream processing platform Apache Kafka\footnote{\href{https://kafka.apache.org}{https://kafka.apache.org}}. In general, no parameters between event logs were altered in the evaluation unless otherwise specified (such as the species definition). Our program interacts with the Kafka cluster via the Faust\footnote{\href{https://github.com/robinhood/faust}{https://github.com/robinhood/faust}} library. We used logs from a public repository containing synthetic event logs \cite{9573336}. Additionally, we used CDLG, a tool for generating event logs with concept drifts \cite{Grimm2022CDLGAT}. The selected event logs vary in complexity in terms of variability, uniqueness of events and length of traces. The event logs are described in more detail in~\Cref{tab:datasets}. To evaluate the completeness- and coverage-based windowing and the used species definition, we set up an event streaming environment.

\begin{table}[ht]
\caption{Simulated Concept Drift Logs}
\label{tab:datasets}
\centering
\begin{tabular}{ l c c c }
\toprule
\textbf{Log Name} & \textbf{\pbox{3cm}{Mean \\ Trace Length}} & \textbf{\pbox{3cm}{Number of \\Events}} & \textbf{\pbox{3cm}{Drift \\ Position}}  \\
\midrule
\multicolumn{4}{ c }{\textbf{Sudden Drift}} \\
\midrule
\makecell{cm-10000 \cite{9573336}} & 25.08 & 125413 & \makecell{Always after 10\%}  \\
\makecell{cd-10000 \cite{9573336}} & 25.52 & 127621 & \makecell{Always after 10\%}  \\
\makecell{cb-10000 \cite{9573336}} & 25.09 & 125447 & \makecell{Always after 10\%}  \\
\midrule
\multicolumn{4}{ c }{\textbf{Recurring Drift}} \\
\midrule
\makecell{Recurring \cite{Grimm2022CDLGAT}} & 12.31 & 61536 & \makecell{Three seasonal changes}  \\
\midrule
\multicolumn{4}{ c }{\textbf{Gradual Drift}} \\
\midrule
\makecell{Gradual 5000 traces \cite{Grimm2022CDLGAT}} & 9.64 & 48203  &  \makecell{Between \\ 40\% and 60\%}  \\
\midrule
\multicolumn{4}{ c }{\textbf{Incremental Drift}} \\
\midrule
\makecell{Incremental \cite{Grimm2022CDLGAT}} & 11.37 & 79608 & \makecell{Five increments}  \\
\bottomrule
\end{tabular}
\end{table}

All event logs in \Cref{tab:datasets} were used in the drift robustness analysis (\Cref{subsec:robustness_to_concept_drifts}) with 1-gram for the species ($\zeta_{act}$), while the real-world logs (RTFMP \cite{RTFMP}, Sepsis Cases \cite{Sepsis}, BPI Challenge logs \cite{BPI-C-2013, BPI-C-2015, BPI-C-2017}, hospital-billing \cite{HB}) were used for accuracy evaluation (\Cref{subsec:accuracy}) relying on all species described in \Cref{subsec:species_def}. The performance evaluation (\Cref{subsec:computational_efficency}) was conducted using the Sepsis Cases log due to its moderate size and complexity.

In particular, we intend to answer the following questions which are derived from the weaknesses of classic windowing methods: \textbf{(R01)} Can the window dynamically adjust to concept drifts without being rigid? \textbf{(R02)} Can we find an accurate window with only minor process knowledge? \textbf{(R03) }How does the approach perform in terms of computational efficiency?

First, we evaluate our approach on datasets with concept drifts (\Cref{subsec:robustness_to_concept_drifts}). Second, we benchmark our approach against different species definitions (\Cref{subsec:accuracy}). Finally, we evaluate the runtime efficiency (\Cref{subsec:computational_efficency}). %
\subsection{Robustness to Concept Drifts}
\label{subsec:robustness_to_concept_drifts}
\begin{figure}[htb]
    \centering
    \includegraphics[width=1\linewidth]{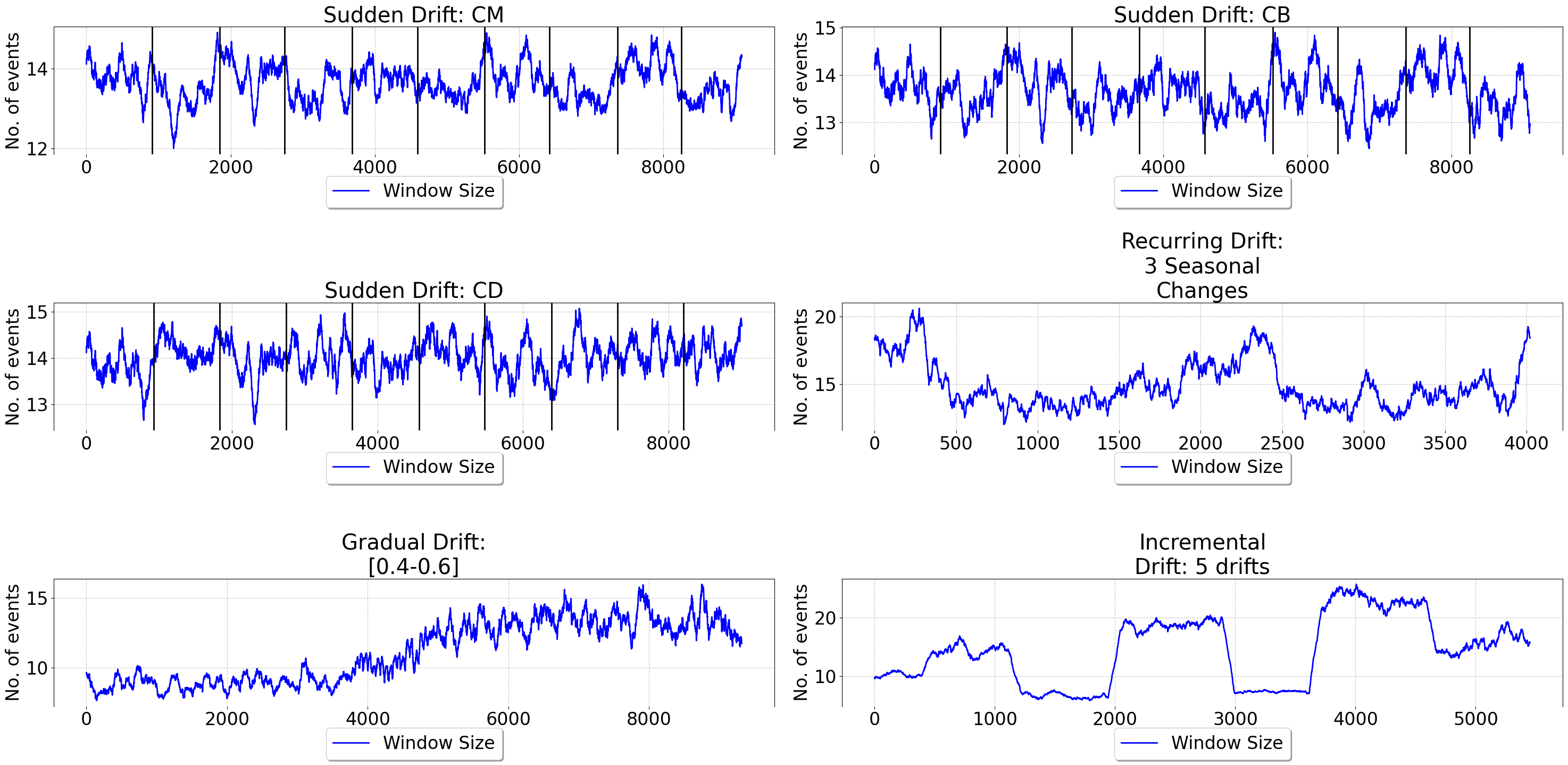}
    \caption{The x-axis shows all windows found for the dataset, while the y-axis shows the number of events captured in the specific window. This demonstrates the adaption of the window to drifts.}
    \label{fig:concept_drift_eval}
\end{figure}

\Cref{fig:concept_drift_eval} shows the results of our experiments on concept drifts and window rigidity when closing a window, considering $\zeta_{act}$ as species definition. With each closing window, we save the size of the window and plot it against the total number of all windows. Here, for all logs (described in \Cref{tab:datasets}), we can see the dynamic adjustment of the window size. 

In fact, the size of the window correlates loosely to the mean trace length, i.e. the window size adapts to the complexity of the process. For instance, a simple process requires less events to be considered complete, while the window for a more complex process increases, capturing the process behavior. This behavior can be observed in the gradual drift log. It starts with window sizes close two 10. When the process gradual drifts into a more complex process, the window size increases to capture the relevant behavior. 

In the case of the sudden drifts, the window detects the change and adjusts the window size accordingly. The sudden drifts are marked with a vertical line. 

The recurring drift demonstrates the adaption to seasonal changes. The windows capture the expected number of events. Gradual and incremental drifts pose a more challenging scenario, where the process changes progressively over a long time. First, the gradual drift shows that if the events vary too much, the window sizes increase to reach coverage. Additionally, if the event variability is too high, a larger window size supports the inclusion of more events, capturing the broader range of activities. Finally, the incremental drift emphasizes this behavior. Due to the complexity of the processes, the window size increases and adjusts with the next incremental shift.

To answer \textbf{R01}, we analyzed window size changes around known drift points. To accomplish this, we calculate the Mean Relative Change, the Standard Deviation of Relative Change, and the Coefficient of Variation. We analyze 10 windows before the drift and 20 windows after the drift for the Sudden Drift: CM and Gradual Drift event logs.   
\textbf{Sudden Drift: CM}: The results show the window size typically adapts by 8.5\% (±4.3\%) during drift periods. A coefficient of variation of 0.269 (0 indicates rigidness and 1 indicates volatile) demonstrates that the approach avoids excessive fluctuations. 
\textbf{Gradual Drift}: The window size gradually increased from an average of 9.3 events pre-drift to 11.3 events during drift, and finally to 13.3 events post-drift. This represents a 21.6\% relative increase during the drift period. A low coefficient of variation during drift (0.113) and a small average trend (0.003 ± 0.111) demonstrated the controlled adoption. These metrics confirm that the approach can dynamically adjust to concept drifts without being overly rigid or volatile.

\subsection{Species Evaluation and Accuracy}
\label{subsec:accuracy}
\begin{figure}[htb]
\begin{minipage}{0.42\textwidth}
\centering
\begin{table}[H]
\caption{Comparison of best/worst species with F1-scores}
\label{tab:species_comparison}
\scriptsize
\begin{tabular}{ c c c }
\toprule
\makecell{\textbf{Log} \\ \textbf{Name}} & \makecell{\textbf{$\zeta$ (Highest)} -\\ \textbf{F1-score}} & \makecell{\textbf{$\zeta$ (Lowest)} -\\ \textbf{F1-score}} \\ 
\midrule
\makecell{Road-Traffic-Fine\\ Mngmt Process} & \makecell{$\zeta_{act}$(5-gram) \\ (0.9257)} & \makecell{$\zeta_{act}$(3-gram) \\ (0.6212)} \\ 
\makecell{Sepsis Cases} & \makecell{$\zeta_{df}$\\ (0.7686)} & \makecell{$\zeta_{tv}$ \\ (0.6282)} \\ 
\makecell{bpi-c-2012} & \makecell{$\zeta_{tv}$ \\ (0.5850)} & \makecell{$\zeta_{act}$(5-gram) \\ (0.3525)} \\ 
\makecell{bpi-c-2013} & \makecell{$\zeta_{act}$(5-gram) \\ (0.9718)} & \makecell{$\zeta_{df}$ \\ (0)} \\ 
\makecell{bpi-c-2015} & \makecell{$\zeta_{act}$(1-gram) \\ (0.6596)} & \makecell{$\zeta_{tv}$ \\ (0)} \\ 
\makecell{bpi-c-2017} & \makecell{$\zeta_{act}$(1-gram) \\ (0.6105)} & \makecell{$\zeta_{act}$(5-gram) \\ (0.4321)} \\ 
\makecell{hospital-billing} & \makecell{$\zeta_{tv}$ \\ (0.9237)} & \makecell{$\zeta_{act}$(4-gram) \\ (0.7170)} \\ 
\makecell{synthetic, \\ online order} & \makecell{$\zeta_{tv}$ \\ (0.9756)} & \makecell{$\zeta_{df}$ \\(exponential) \\ (0.6709)} \\ 
\bottomrule
\end{tabular}
\end{table}
\end{minipage}
\hfill
\begin{minipage}{0.5\textwidth}
\centering
\includegraphics[width=\linewidth,height=0.45\textheight]{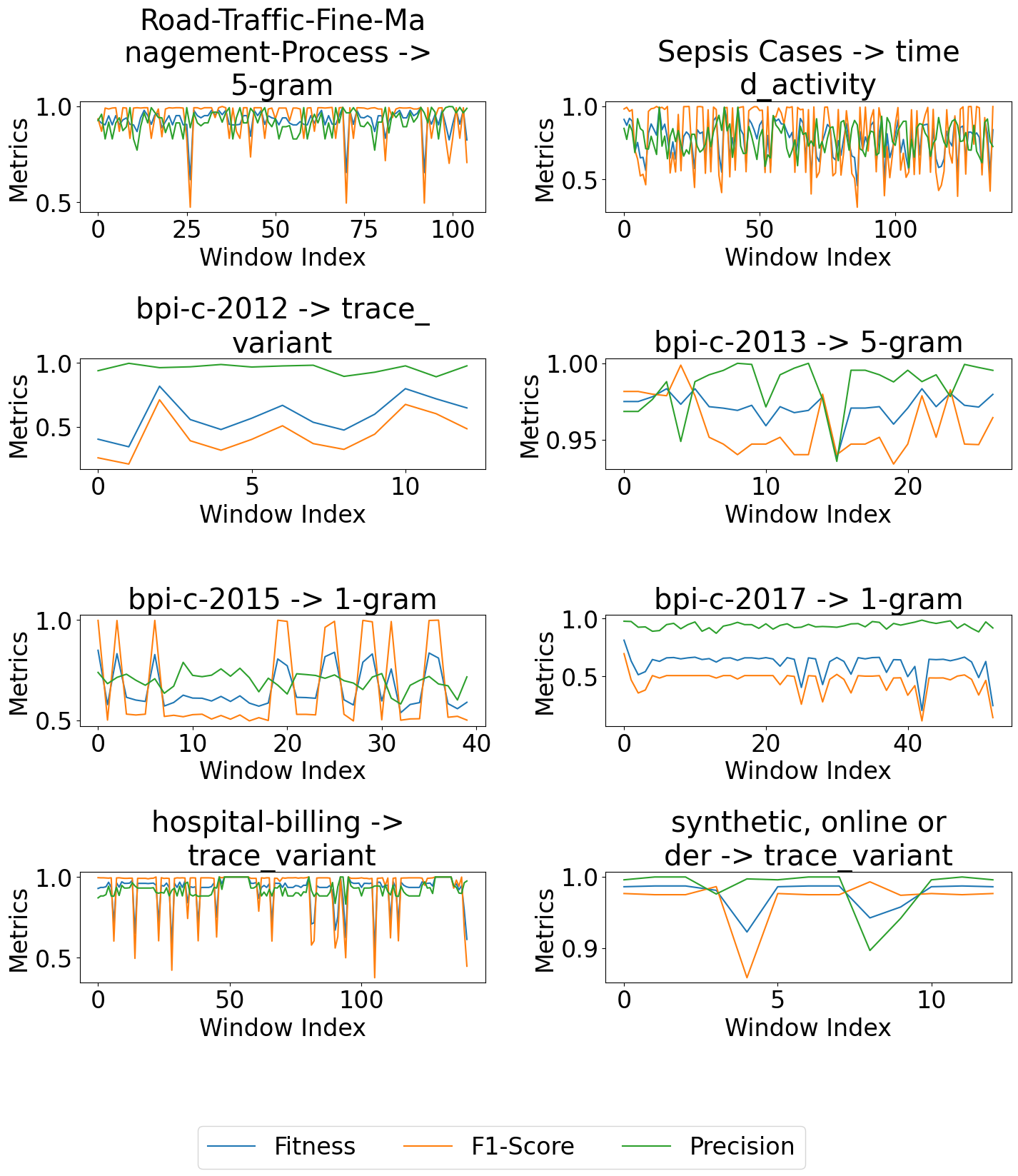}
\caption{Conformance metrics for all event logs over all windows}
\label{fig:f1_over_time}
\end{minipage}
\end{figure}

Next, we assess the impact of the species definitions on coverage estimation. We conducted experiments by sequentially applying different species retrieval functions to the event stream. Specifically, we processed the first 2,000 events of each event log. Then, we applied the species retrieval functions to the coverage estimation, and recorded conformance metrics (i.e., including Fitness, Precision, F1-score) once a window was identified. \Cref{tab:species_comparison} summarizes our findings.

It presents the species retrieval function that achieves the highest and lowest average F1-score across all windows and various event logs. The $\zeta_{act}$ performs particularly well for logs with structured and repetitive sequences, as it captures local activity patterns (e.g., bpi-c-2013). In contrast, $\zeta_{df}$ species definitions excel in logs where the timing between events is crucial (e.g., Sepsis Cases). $\zeta_{tv}$ species are more suited for highly variable, complex logs, where the diversity in process behavior is more pronounced (e.g., hospital-billing). 
Additionally, \Cref{fig:f1_over_time} illustrates the evolution of the conformance metrics over all windows. For instance, in the case of \emph{bpi-c-2013}, the F1-score remains consistently high, supported by an average window size of 69.44 (Min: 25, Max: 187). By contrast, the \emph{Sepsis Cases} data exhibits significant variability, with an average window size of 14.53 (Min: 4, Max: 118). This volatility may be attributed to the differences in process execution patterns, possibly due to varying practices during night and day shifts. Furthermore, the \emph{Road-Traffic-Fine-Management-Process} reveals three notable precision drops. Here, the average window size is 18.96 (Min: 10, Max: 72). In most cases, windows begin with the activity \emph{Create Fine} and conclude with either \emph{Send for Credit Collection} or \emph{Payment}, which aligns with the expected process flow. For those instances where precision falls, the window instead starts with \emph{Payment}, potentially indicating deviations in process execution.

To answer \textbf{R02}, we compare the average F1-score of the best species with a fitting windowing approach. In particular, we utilize the Sepsis Cases Event Log with a Landmark Window and choose the activity label \emph{ER Registration}. Please note that, this assumes that the process has a clearly defined starting activity. As the sepsis cases fulfil this, the landmark window will therefore achieve the best results, even if it is susceptible to concept drifts. The average F1-Score over all discovered windows is 0.62  (min: 0.58, max: 0.88). Further, a count-based tumbling window with a fixed size of 20 achieves an average F1-Score of 0.61 (min: 0.34m max: 0.81). Both archive significant lower F1-Scores compared to our approach (0.76). A benchmarking of the other windows and event logs is available in the repository.

\subsection{Performance Evaluation}
\label{subsec:computational_efficency}
\begin{figure}[H]
    \centering
    \includegraphics[width=1\linewidth]{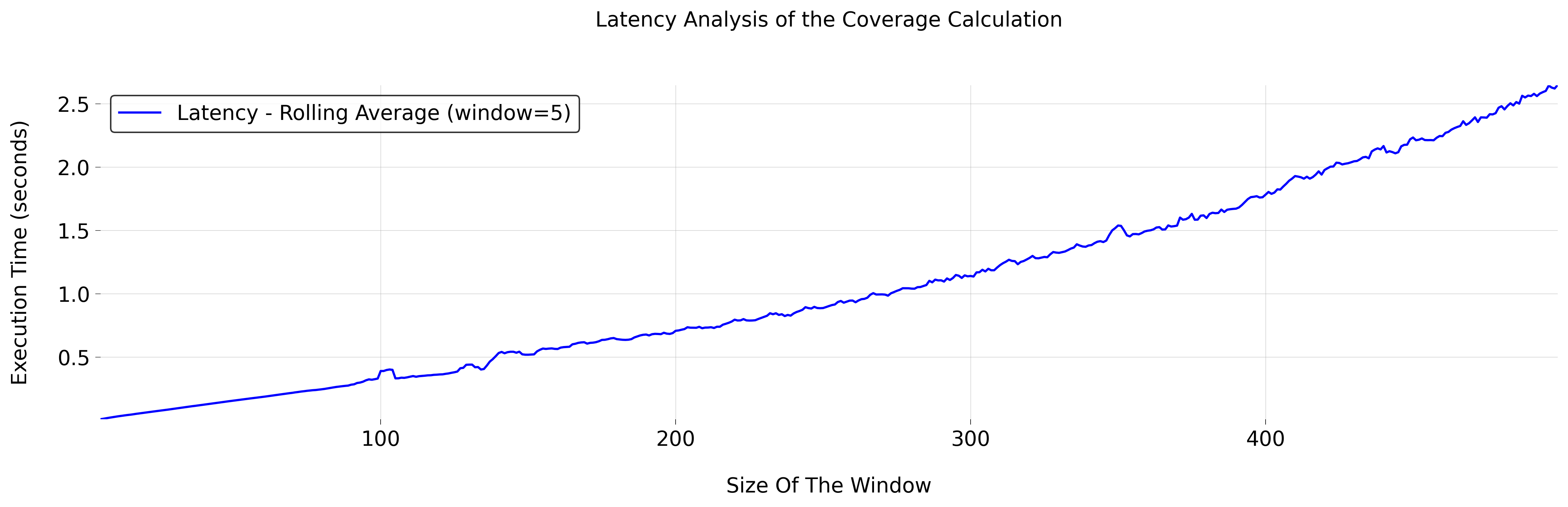}
    \caption{Analysis of the latency time for incrementally larger windows ($n_{\text{max}} = 500$)}
    \label{fig:performance-evaluation}
\end{figure}

Now, we report on evaluation results in terms of runtime efficiency.
We evaluate latency and throughput at the system level, which are considered conventional metrics for assessing the efficiency of stream data processing systems~\cite{Karimov2018}. First, we define latency as processing time latency, i.e., the time interval between the ingestion of the first event within a window and the emission of the result. To evaluate latency, we use the python library perfplot.\footnote{\href{https://github.com/nschloe/perfplot}{https://github.com/nschloe/perfplot}} As shown in \Cref{fig:performance-evaluation}, the results confirm our assessment of time complexity (see \Cref{subsec:completeness_streaming}), i.e. we observe a linear (O($S_N$)) increase in latency.

To answer \textbf{R03}, we use the Throughput. It is defined as the number of events that are processed in a given time interval. That is, the unit of throughput is processed events per second. We calculate an average throughput of 8962.13 events/s, with a standard deviation of 1166.18 events/s. However, external factors such as the Kafka Cluster and the Faust implementation influence this result. Notably, since Faust is implemented in Python, there might be a decrease of performance.

The results show that our approach can find accurate windows only with minor process knowledge. Furthermore, It can dynamically adjust to concept drifts without being rigid, while, being efficient in terms of utilizing computational resources.

\section{Related Work}
\label{sec:related_work}
Recently, several proposals for process mining over event streams have been made~\cite{burattin_streaming_2022}. Typically, streaming algorithms maintain some computational state based on the properties of
past events, which may employ approximations, such as lossy counting~\cite{DBLP:conf/cec/BurattinSA14}. 
The computational state may be managed separately, which then enables the integration of offline discovery algorithms~\cite{DBLP:journals/kais/ZelstDA18}. Streaming algorithms have also been presented to assess the conformance of event streams with a process model, e.g., by evaluating model constraints over a stream~\cite{DBLP:conf/bpm/BurattinZADC18}, by replaying the events from the stream in the model~\cite{DBLP:conf/bpm/BurattinC17}, by aligning trace prefixes~\cite{DBLP:journals/ijdsa/ZelstBHDA19}, or by learning a state predictor~\cite{DBLP:journals/is/LeeBMS21}.

The above techniques ignore windows over streams (e.g., by assuming that events are partitioned in traces and traces will end eventually) or implement simple heuristics to select the window size, see \Cref{sec:window_size}. In this work, we propose a systematic approach to dynamically select an appropriate window size. \\
The question of how to determine a suitable window size is closely linked to sample-based process mining. However, these measures and existing sampling strategies are typically defined for a fixed population of events, which does not materialize in a streaming setting~\cite{bauer2018much,bauer2022sampling}.

Our selection of a window size is guided by completeness estimators from biodiversity research. In addition to the mentioned Chao2-estimator, species richness may also be assessed using other estimators, such as abundance-based coverage~\cite{chao1992estimating}. Those estimators, as well as estimators for further population properties~\cite{chao2014rarefaction}, provide an avenue for further research on window selection mechanism in process mining.

\section{Conclusion and Future Work}
\label{sec:conclusion}
This paper presents a novel approach for streaming process mining relying on species discovery and dynamically adjusted sliding windows aiming to determine the optimal window size to preserve the accuracy and completeness of the discovered process models. The comparison of our approach with sliding window approaches utilizing time-based, count-based and landmark windows show superiority in terms of accuracy and robustness to concept drifts.
Future work shall address the transferability of our approach to additional process mining scenarios involving both highly fluctuating time intervals and high variability of process changes. Furthermore, the application of the proposed approach in distributed scenarios (i.e., dealing with volume, velocity and variability) is planned, where scalability as well resource constraints are essential.
Additionally, we plan to consider behavioral aspects like resource allocation and workload distribution. 

\medskip
\textbf{Acknowledgements.} This work was partly supported by the German Federal Ministry of Education and Research (BMBF), grant number 16DII133 (Weizenbaum-Institute). This work received funding by the Deutsche Forschungsgemeinschaft (DFG), grant 496119880. The responsibility for the content of this publication remains with the authors.

\printbibliography

\end{document}